\documentclass[showpacs,amsmath,amssymb,twocolumn,aps]{revtex4}

 \usepackage{graphicx}
\usepackage{epsfig}
 \usepackage{pstricks}
 \usepackage{pst-plot}
  \usepackage{psfrag}
 \usepackage{dcolumn}
 \usepackage{bm}

 \def\comment#1{}
 \def\mn#1{}

\def\lfrac#1#2{#1/#2}

\begin{document}

\title{High Bias Voltage Transport in Metallic Single-walled Carbon Nanotubes 
under Axial Stress}

\author{J\"urgen Dietel}
\affiliation{Institut f\"ur Theoretische Physik,
Freie Universit\"at Berlin, Arnimallee 14, D-14195 Berlin, Germany}
\author{Hagen Kleinert}
\affiliation{Institut f\"ur Theoretische Physik,
 Freie Universit\"at Berlin, Arnimallee 14, D-14195 Berlin, Germany}
\affiliation{ICRANeT, Piazzale della Repubblica 1, 10 -65122, Pescara, Italy}
\date{Received \today}

\begin{abstract}
We calculate the current-voltage characteristic of 
a homogeneously  strained metallic carbon nanotube adsorbed on a substrate. 
The strain generates a gap 
in the energy spectrum leading to a 
reduction of the current. 
In the elastic regime, the current-voltage characteristic 
shows a large negative differential conductance at bias 
voltages of around $ \gtrsim  0.17 $V.
We discuss the implications for the current in the superelongated regime.  
\end{abstract}


\maketitle

\section{Introduction} 
Carbon nanotubes have many electronic and mechanical applications. 
Their ability
to sustain high currents before breaking  
was observed
in a number of experiments 
on  
metallic single-walled nanotubes (SWNT) 
adsorbed on a substrate
\cite{Yao1, Park1, Javey1}.
Corresponding theoretical 
work was done in Refs.~\onlinecite{Lazzeri1, Lazzeri2}.   
With respect to the mechanical 
properties, 
carbon nanotubes belong 
to the strongest and stiffest materials discovered so far.
Under stress, a tube  will 
first expand elastically.
After reaching the yield point, 
it undergoes plastic deformations, leaving a permanent deformation.
This point was measured at strains around $ 5-10\% $ 
for nanotubes with diameter $1-3$nm depending on the 
chirality \cite{Zhang1, Bozovic1}.       

When the effective temperature of the nanotube 
caused by the intense electron-phonon scattering at a high bias voltage
(we assume in this paper that the system is not externally heated) exceeds
the  activation barriers of defects,
large spiral-like defect stripes are created along the tube \cite{Huang1}.
These defect 
stripes correspond to missing lattice plane pieces in the crystal. 
The missing atoms of these plane pieces are rearranged 
at the end points of the nanotube in such a way  
that a  superelongation of SWNTs of up to $ 270\% $ before 
cracking is reached. 
In Ref.~\onlinecite{Dietel1}, we
have developed
a theory for this effect in a simplified
crystal model.
Finally, when the effective temperature due to the applied bias voltage 
is lower than the activation barrier 
the system responds to the external stress 
with large non-homogeneous defect pile ups or 
immediately bond-breaking depending on the physical surrounding 
\cite{Yakobson1, Nardelli1}. This  leads then to necking of the SWNT.
Here, we will not address the current-voltage characteristic in this regime.

The purpose of this paper is to analyse the 
electrical properties in the stress ranges where nanotubes 
show a homogeneous behavior in space, 
i.e., in the elastic regime  of 
small elongation, and in the superelongation regime. 
We will first discuss the elastic regime, 
and the superelongation regime will be discussed at the end. 
\mn{Der vorige Paragraph is frustrating:
You say: the high T regime is discussed at the end.
The low T regime is not discussed at all in this paper.
Was gibt es sonst noch für regimes.
 }

If subjected to an external homogeneous stress, 
a nanotube responds with 
a homogeneous strain. This results in modifications 
in the bond lengths and leads, via
the electron-phonon interaction, to a gap in 
the electron spectrum of metallic SWNTs. This
strongly modifies the current-voltage characteristic, and 
represents, therefore, a  
promising effect  
for building sensitive stress sensors.  

On a 
substrate, the current increases 
with voltage,
whereas  for suspended SWNTs it decreases, i.e., it shows
a small negative differential conductance at high voltages \cite{Pop1}.
Similarly, we will find here a negative differential 
conductance which is now very large for a SWNT adsorbed on a 
substrate under external stress at bias 
voltages of around $ 0.17 $V\mn{how much?}.

 The energy levels of electrons in a nanotube form
one-dimensional 
bands in the graphene Brillouin zone around the $ {\bf K} $  and 
 $ {\bf K}' $  points. For metallic SWNTs, 
there are two energy bands 
corresponding to right- and left-moving electrons 
which
cross at these 
points. 
We assume here that the diameter of the nanotube is so small 
that 
electron excitations to higher bands
are negligible. For example this approximation is valid at bias voltages   
$ U \lesssim 2 $V for nanotubes with diameter $ 2$nm. 
These diameters are typical in present day current-voltage 
experiments \cite{Yao1, Park1, Javey1}. 
But we expect the validity of the approximation even 
beyond this value since electron 
scattering to higher bands is effectively forward 
due to the band edges where electrons reverse their direction. 
Note that we use in our numerical calculations below scattering values  
for nanotubes with a diameter of $ 2 $nm \cite{Dietel1}. 

In order to calculate the current-voltage characteristic 
at high bias voltage we will 
use the semi-classical Boltzmann equation for electrons and phonons. 
The various scattering mechanisms between electrons, phonons 
and impurities will be discussed first in Sect. II. Then we
will discuss in Sect. III the effect of strain on the wavefunctions and 
the energy spectrum of the electrons in the nanotube. This leads to 
modifications in the electron-phonon scattering times. Section IV 
introduces the different Boltzmann equations for the electrons and 
phonons which we use in order to calculate the current-voltage 
characteristic. The results of this calculation will be discuss  
in Sect. V ending up with a conclusion in Sect. VI.

\section{Scattering Processes} 

The method we use here to calculate 
the current-voltage characteristic of metallic nanotubes is based on 
the semi-classical Boltzmann equation.
Within this method quantum 
interference corrections to the conductance are not taken into account 
\cite{Datta1}. It was recently shown numerically 
that this correction to the conductivity 
is negligible above room temperature for single-walled 
carbon nanotubes without structural defects 
due to phonon scattering decoherence mechanisms \cite{Ishii1}.
This temperature is immediately reached through electron-optical phonon  
and further optical phonon-acoustic phonon scattering in  
nanotubes under  high bias voltage.      
 
At low voltages innervalley  elastic scattering, which consists of 
 acoustic phonon 
scattering (in the quasi-elastic limit) 
and impurity scattering, is most relevant.
We neglect electron-electron collisions. 
At higher voltages inelastic scattering between the electrons 
and optical phonons, which then leads to intervalley 
scattering, becomes relevant. 
This scattering process leads at high bias voltage to the creation of 
hot optical phonons, which are then no longer in thermal equilibrium with 
the environment \cite{Lazzeri2}. 
This leads us to the conclusion that we have   
to consider besides the electrons also phonons  
in the  semi-classical Boltzmann approach.      
The hopping between an electron in the  
$ {\bf K} $ valley and the other at the $ {\bf K'}$-valley  
 is  mediated by zone-boundary 
optical $ A'_1 $ $ {\bf K} $-phonons where only Kekul\'{e} 
type of lattice distortions
couple to the electronic system \cite{Suzuura1}.
On the other hand  jumps of electrons within the same valley are mediated 
by zone-center $ E_{2g} $ $ {\bf \Gamma} $-phonons \cite{Lazzeri1}.   
The optical phonons decay to lower lying secondary acoustic phonons 
by phonon-phonon scattering \cite{Dietel3}. These then decay to other 
acoustic phonons or phonons in the substrate and to a smaller amount also 
in the leads. We show in Fig.~1 this electron-phonon relaxation path.

\begin{figure}
\begin{center}
\includegraphics[clip,height=7.5cm,width=8.5cm]{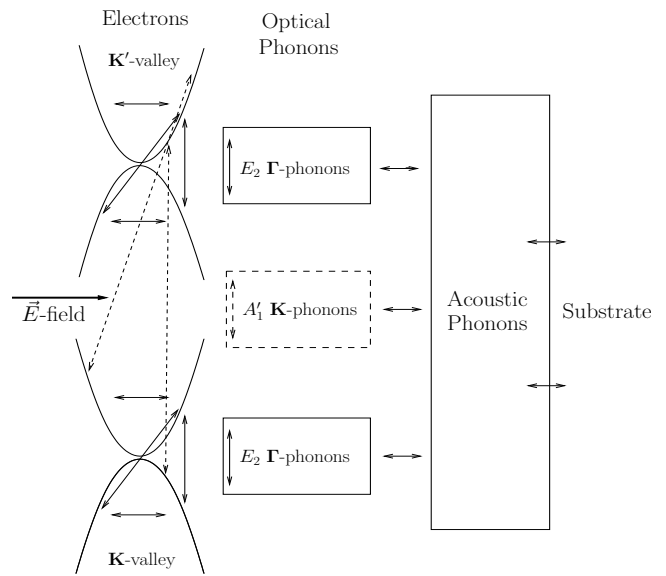}
\vspace*{-0.2cm}
\caption{Main electron-phonon relaxation 
path of metallic SWNTs at high bias voltage 
where electron-optical phonon scattering is dominant. Innervalley 
electron jumps shown as solid  
doublearrows in the figure are mediated by 
$ E_2 $ $ {\bf \Gamma} $-phonons. Intervalley electron jumps  mediated by 
$ A'_1$ $ {\bf K} $-phonons are shown by 
dashed doublearrows.}
\end{center}
\end{figure}

\section{Strain Effects on the electron system}

By using the electron-phonon 
interaction Hamiltonian for
acoustic phonons as in 
Ref.~\onlinecite{Suzuura1}
we have calculated the 
lowest-band eigenfunctions and 
eigenvalues for a metallic nanotube under stress.
For electrons in the $ {\bf K} $ valley, 
these are given by 
(letting the axis of the SWNT point in the $y$-direction)
\begin{align} 
&\!\! \Psi^K_{s,k}(y) = e ^{i y (k + k_{y_0})}        
\genfrac(){0pt}{0}{\frac{ k_-}{|k_-|}} 
 { s}    ,  \label{300}   
\end{align}  
where $ s=1 $ for the conduction band , $ s=-1 $ for the 
valence band and $ k $ is the momentum in $ y $-direction. 
Moreover, we used the abbreviation 
 $k_\pm\equiv k_{x_0}\pm  i k $
according to 
\begin{equation} 
k_{x_0}+ ik_{y_0}= \frac{g_2}{\hbar v_F} e^{i 3 \eta}  
 \left[ (u_{xx}\!-\! u_{yy}) + i  2  
    u_{xy} \right].
  \label{310} \end{equation} 
Here $ v_F $ is
the electron velocity, 
$ u_{ij}$ the strain tensor,
 and
 $ g_2 \sim 2  $eV \cite{Suzuura1}  
the electron-phonon 
coupling constant.
The parameter $ \eta $ 
in $k_{x_0,y_0}$ is the chiral angle of the SWNT.  
  For electrons in the  $ {\bf K}' $
valley we obtain $ \Psi^K_{s,k}(y) $ with the 
substitution $ g_2 \rightarrow -g_2 $ and $k_{x_0}-i k 
\rightarrow k_{x_0}+i k $.

The energy eigenvalues are 
\begin{equation} 
\epsilon(s,k)= s \hbar v_F \sqrt{k^2_{x_0}+ k^2} ,
                  \label{320}  
\end{equation} 
Strain makes
$ k_{x_0} \not=0  $ and opens
a gap
$ \Delta= 2 \hbar v_F |k_{x_0}| $
 in the electron spectrum
of a formerly metallic SWNT. This is
consistent with numerical calculations and experiments 
\cite{Yang1, Minot1, Cao1, Filho1, MHuang1}. 
This spontaneous gap generation is {\em not\/} 
seen by computer calculations
in graphene \cite{Pellegrino1}.     
The energy eigenvalues $\epsilon(s,k) $ 
show an important exception: 
The spectrum remains gapless \cite{Yang2, Maiti1} 
for a metallic SWNT  with $ \eta=\pi/6 $ known as armchair nanotube,
where the homogeneous axial stress obeys  $ u_{xy} =0 $.
From this and the discussion below it is clear that armchair nanotubes
show only a small effect on the current-voltage 
characteristic by an applied  external stress.  
For zigzag SWNTs, for which one has $ \eta=0 $,  
we obtain the maximal gap value 
for a given applied axial external stress.    

Let us now discuss the current-voltage characteristic due to strain. 
For a formerly metallic nanotube, we obtain that the energy band  
and the electron-phonon scattering times are modified. Indeed, 
both depend effectively  on 
the gap parameter $ \Delta $ which is itself a function  
of the strain and the chiral angle of the nanotube. 
We neglect here 
modifications of the eigenfunctions above due to $ k_{y_0}\not=0 $ 
as this leads effectively only to small momentum shifts of the zone 
boundary phonon velocities.
The corresponding modifications of the current-voltage characteristic
are small due to the fact that the effective phonon 
mean free path is much smaller than the SWNT length.
This has been checked numerically. 
We now discuss the modifications of the effective electron-phonon 
scattering times due to the  
wavefunction modifications under stress.       
     
We restrict ourselves
first to the $ {\bf K} $ 
valley and zone-center phonon scattering.
The  results for intervalley scattering are mentioned below Eq.~(\ref{45}). 
In contrast to the current-voltage
characteristic of SWNTs without band gap \cite{Lazzeri2} we
have to take backward scattering (inner and interband) and interband 
forward scattering into account.   
Interband scattering leads, for example, to the result that
the conduction band can be filled with electrons 
by phonon absorption or depleted by phonon emission.  
Therefore, we should take into account that we have from the beginning 
electrons (holes) injected in the conduction 
(valence) band in the SWNT by a non-zero lead temperature. 
Without interband scattering, only these electrons 
and the corresponding holes contribute 
to the current (see the accompanying discussion to Eq. (\ref{50})).

The electronic zone-center phonon scattering Hamiltonian is given by 
\cite{Ishikawa1} 
\begin{equation}    
H_{\rm int}=-\frac{g_3}a  \left( \begin{array} { c c } 
0 & u_y({\bf r}) + i u_x({\bf r}) \\
u_y({\bf r}) - i u_x({\bf r}) & 0 
 \end{array} 
\right) ,                              \label{330}
\end{equation} 
where $ a $ is the nearest-neighbour distance in the  
lattice and $g_3$
was determined in 
Ref. [\onlinecite{Lazzeri1}] by density functional methods 
for metallic SWNTs. 

Next we calculate the
momentum-dependent electron-phonon transition 
matrix elements $ |\langle \Psi^K_{s, k}|
H_{\rm int} | \Psi^K_{s', k'} \rangle|^2 $, which are proportional   
to the effective electron-phonon scattering times. 
In the following, we mention for which type of phonons, i.e. longitudinal 
($ u_y \not=0 $ and $ u_x=0 $) or transversal ($ u_y =0 $ and $ u_x \not= 0 $), 
the matrix element  $ |\langle \Psi^K_{s, k}|
H_{\rm int} | \Psi^K_{s', k'} \rangle| $ is non-vanishing:  
\begin{eqnarray}
&&
 \!\!\! \!\!\! \!\!\!\!\!
|\langle \Psi^K_{s, k>0}| H_{\rm int} | \Psi^K_{s', k'<0 }\rangle|^2  
\!\!\propto  \!
 \left|
\lfrac{k_+}{|k_+|}
 \!+\! \lfrac{k'_-}{|k'_-|}
\right|^2 
 \approx  4         ,   \label{340}    \\
& &
 |\langle \Psi^K_{s>0, k\gtrless 0}| H_{\rm int} | \Psi^K_{s'<0, k'
 \gtrless0} \rangle|^2
 \approx  4   \,.      
\label{350}  
\label{@}\end{eqnarray}
The matrix element
(\ref{340})
comes from interband forward scattering, which means $ s \not= s'$, 
with
transverse  optical phonons 
($ u_x \not =0 $).  For innerband 
backward scattering, i.e. $ s=s'$, this non-zero matrix element 
is mediated by longitudinal optical phonons ($ u_y \not=0 $). 
The non-zero matrix element 
(\ref{350}) 
come from interband backward scattering
with longitudinal optical 
phonons ($ u_y \not= 0 $) 
if the gap is small ($ |k_{x_0}| \ll |k|, |k'|$),
and from 
transversal optical phonons ($u_x \not= 0 $) 
if the gap is large   
($ |k|, |k'| \ll |k_{x_0}| $). 
Note 
that (\ref{340}) and (\ref{350}) 
become
exact for large and small gaps and
are approximate 
for intermediate gap values. 
Note also that in the present approximation,
 (\ref{340}), (\ref{350}), the 
backward scattering matrix elements do not depend on 
the stress, i.e., we can set $ k_{x_0} = 0 $.  

\section{Current-Voltage Characteristic} 
In section II we introduced a two-valley model for 
the scattering mechanisms of the SWNT under bias voltage.   
In the following, we use the approximation of an 
effective one-valley model 
coupled to only one sort of optical phonons. 
In fact, this was already used by us   
in Ref.~[\onlinecite{Dietel3}] to calculate analytically 
the current-voltage characteristic 
 of unstrained SWNTs with non-reflecting leads. 
This is a simplification of the two-valley 
Boltzmann approach where scattering between the valleys is mediated by 
zone-boundary and zone-center optical phonons 
\cite{Lazzeri2, Dietel3} which we discussed in Sect.~II.
The latter  model leads to current-voltage characteristic  of unstrained 
SWNTs \cite{Lazzeri2, Dietel3} which are in good accordance with experiments. 
Our one-valley approximation is justified by the fact that 
the optical frequencies of the zone-center phonons and zone-boundary phonons, 
and also the electron-phonon coupling in the phononic sectors are 
rather similar. The electron-optical phonon interaction in this model leads to
an effective inverse scattering time $ 1/\tau_{\rm ep} $ 
for the electrons and $ 
2 s^p/\tau_{\rm ep} $ for the phonons with parameters found in 
Ref.~\onlinecite{Dietel3}.   

We denote by $ f_{L}(k,x,t) $ ($ f_{R}(k,x,t) $) 
the left (right) moving electron distribution functions, and
by $ n^{fs }(k,x,t), ~ n^{bs }(k,x,t) $
the distribution function of 
the phonons mediating forward or backward scattering 
of electrons, respectively.
The time evolution of the electrons is governed by the 
semi-classical Boltzmann-equation 
\vspace{-2.2mm}
\begin{equation}
\left[\partial_t \mp v_e(k) \partial_x +\frac{e {\rm E}}{\hbar} 
  \partial_k\right]
f_{L/R}= \left[\partial_t f_{L/R}\right]_c \,.       \label{10}
\end{equation} 
~\\[-1.2em]
with $ v_e(k)= |\partial \epsilon(s,k)/\partial (\hbar k) | $,   
$ [\partial_t f_{L/R}]_c \approx [\partial_t f_{L/R}]_e+ 
+[\partial_t f_{L/R}]_{fs}+ 
[\partial_t f_{L/R}]_{bs} $, and elastic 
scattering 
\begin{equation} 
[\partial_t f_{L}]_e = (v_F/l_e) [f_R(k)-f_L(k)] \,  
\tilde{\nu}(\epsilon(s,k))          \label{15} 
\end{equation} 
consisting of acoustic phonon scattering (in the quasi-elastic limit) 
and impurity scattering where $ l_e $ is the elastic scattering mean free 
path.  
We denote by $ \nu(\epsilon(s, k)) $ the density of states of the left 
or right-moving electrons, i.e. $ \nu(\epsilon(s,k))=
1/|\partial \epsilon(s,k)/\partial k| $ and $ \tilde{\nu}=
\nu(\epsilon) \hbar v_F  $ is the dimensionless density of states.

The time evolution of optical phonons is given by 
\begin{equation}
\left[\partial_t +v_{\rm op}(k) \partial_x\right]
n^{bs / fs}= \left[\partial_t n^{bs/fs}\right]_c +
\left[\partial_t n^{bs/fs} \right]_{\rm osc} \,.         \label{20}
\end{equation}
Here $ v_{\rm op} $ is the optical phonon velocity \cite{Dietel3}. 
$ \left[\partial_t n^{bs/fs} \right]_c $ is 
a phonon-electron  scattering term consisting of the interaction 
with backward or forward scattered electrons.  
$ \left[\partial_t n^{bs/ fs}\right]_{\rm osc} $ is a thermal phonon 
relaxation term such that the coupled electron-phonon system is not heated up 
by applying large voltages on the SWNT.

Scattering of phonons with electrons leads to the following 
scattering contributions
(backward scattering and forward interband scattering)   
in the electronic Boltzmann equation (\ref{10}) and in the phononic 
Boltzmann equation (\ref{20}) (we restrict ourselves  to 
$ k>0 $ phonons in (\ref{40}) for simplicity)   
\begin{widetext}
\begin{align}
&  
\left[\partial_t f_{L}\right]_{bs / fs}(k,x) =  \label{30}  \\
& \frac{1}{\tau_{\rm ep}}
 \bigg( \! \tilde{\nu}(\epsilon^{+} ) r^{bs/fs}(\epsilon^+)
       \bigg\{ \! [n^{bs /fs} (k^+,x)+1]
f_{R / L} (k_{R/ L}(\epsilon^{+}))
[1-f_L(k_L(\epsilon))]   
-n^{bs/ fs}(k^+,x)[1 - f_{R/L} (k_{R/L} (\epsilon^{+}))]  
f_L(k_L(\epsilon))\!  \bigg\}
     \nonumber     \\     
& \! + \! \tilde{\nu}(\epsilon^{-}) r^{bs/fs}(\epsilon^-)  \bigg\{\!  
n^{bs / fs}(-k^-,x)  f_{R / L}(k_{R/ L}(\epsilon^{-})) 
[1\! - \! f_L(k_L(\epsilon))]  
\! - \! [n^{bs/ fs}(-k^-,x)\! +\! 1][\! 1-\! f_{R /L} (k_{R/ L}(\epsilon^{-}))]
f_L(k_L(\epsilon))\, \!  \bigg\} \! \!  \bigg),      \nonumber 
\end{align} 
\begin{align}
& 
\left[\partial_t n^{bs/ fs }\right]_c(k,x)
=2 \frac{ s^{p}}{\tau_{\rm ep}} \,
\left|\frac{1}{\tilde{\nu}(\epsilon(k_L^+))}  
\pm 
\frac{1}{\tilde{\nu}(\epsilon(k_L^-))}\right|^{-1}   
\bigg( [n^{bs/ fs }(k,x)+1]\bigg \{f_{R}(k_{R}^+)  
[1-f_{L/R}(k_{L/R}^-)]      \label{40}   \\
&  +f_{L/R}(-k_{L/R}^-)[1-f_{R}(-k_{R}^+)] \bigg\} -  n^{bs/ fs }(k,x) 
             \bigg\{f_{L/R}(k_{L/R}^-)          
          [1-f_{R}(k_{R}^+)]  +f_{R} (-k_{R}^+)
[1-f_{L/R}(-k_{L/R}^-)]\bigg\} \bigg) \nonumber 
\end{align}  
\end{widetext}
with the abbreviations $ k^\pm=k_R(\epsilon^{\pm})  
-k_L(\epsilon) $, $ \epsilon^\pm= \epsilon \pm \hbar \omega $ in (\ref{30}). 
In (\ref{40}) we used the abbreviation 
\begin{equation}  
 k_{R/L}^{\pm}=\pm \frac{k}{2} + \frac{\omega}{2 v_F}  
\sqrt{\frac{\frac{\omega^2}{v^2_F}-k^2-4 k_{x_0}^2}{\frac{\omega^2}{v^2_F} -k^2}}. \label{45} 
\end{equation}
where the right hand side of Eq.~(\ref{45}) is independent of $ R, L $ 
and $ k $ is the phonon momentum. 
 The numerical constant $ s^p $ is given by 
$ s^p\! \approx\! 0.67 $ \cite{Dietel3}. 
The frequency of the optical phonon is denoted by $\omega $.   
The factors $ r^{bs/fs}(\epsilon^\pm) $ within our one-valley model 
are determined by the influence of zone-boundary $ {\bf K} $-phonon 
 scattering on the 
effective electron-phonon scattering time by carrying out a similar overlap 
calculation for these phonons as was done for zone-center 
$ {\bf \Gamma} $ phonons below 
Eq.~(\ref{350}). Here we use the zone-boundary phonon scattering Hamiltonian in 
Ref.~\onlinecite{Suzuura2} together with similar approximations used below 
Eq.~(\ref{350}). This leads to $ r^{bs}(\epsilon^\pm) \approx 
0.7 (1/\tilde{\nu}(\epsilon^\pm)+ 1/\tilde{\nu}(\epsilon))^2/4 + 0.3 $ 
for backward scattering and $ r^{fs}\approx 0.3  $ 
for interband forward scattering \cite{Dietel3}.  
   
We restrict ourselves here to short SWNTs 
with lengths between $ 50 $nm where phonon scattering 
becomes relevant and $ 500 $nm. 
For the thermal phonon relaxation term
$ \left[\partial_t n^{bs/  fs}\right]_{\rm osc} $ we use 
an expression which takes into account also the second 
generation acoustic phonons with parameters discussed 
in the case of the metallic SWNTs without external stress in 
Ref.~\onlinecite{Dietel3}.
Since $ |1/\tilde{\nu}(\epsilon(k_L^+))- 
1/\tilde{\nu}(\epsilon(k_L^-)) |^{-1} \gg 1 $ in (\ref {40}) we can 
disregard,
 in the calculation of Eq.~(\ref{20}), the term
$ \left[\partial_t n^{fs} \right]_{\rm osc} $,  
and the term
$ v_{\rm op}(k) \partial_x 
n^{fs} $  
for the forward scattering phonons.
To integrate the Boltzmann equations (\ref{10})-(\ref{40}) we use 
a generalisation of a 
discretized  time-splitting method described 
in Ref.~\onlinecite{Dietel3} for the stress-free case. 
Due to the change in the density of states coming 
from the external stress, the momentum and space-grid is no longer
equidistant in position space but still equidistant in momentum space
\cite{REMA}.
The large prefactor $ |1/\tilde{\nu}(\epsilon(k_L^+))- 
1/\tilde{\nu}(\epsilon(k_L^-))|^{-1} $ in (\ref{40}) allows us  
to determine $  \partial_t n^{fs }$ in every iteration process 
by (\ref{40}) with $ \left[\partial_t n^{fs }\right]_c(k,x)=0 $. 
This dynamics leads approximately to the same 
stationary solution of (\ref{10}), (\ref{20}) 
as in the case  when using the full phonon 
scattering dynamics (\ref{40}).

\section{Discussion} 
The upper panel in Fig.~2 with the continuous curves 
shows the current-voltage characteristic
calculated with the help of (\ref{10})-(\ref{40}) for a SWNT
of length $ L= 300 $nm and different values 
of the gap energies $ \Delta $. Note that 
the black circles are calculated for gap energies $ \Delta \to  0$ just after 
the gap opening.
The recorded data points are restricted by the demand that the 
momentum grid, the energy 
values $ \Delta $ and optical phonon frequencies $ \omega $ are commensurate. 
By comparing the current values $ I $ for voltage $ U =1.5 $V and 
very small gap energy $ \Delta \to 0 $ with the 
corresponding values for metallic SWNTs \cite{Lazzeri2, Dietel3} 
we obtain that the current values are around a factor six times smaller 
after the gap is opened. The crosses at the curves in Fig.~2 
shown for $ \Delta < \hbar \omega $ are calculated using the full 
two electron valley model coupled to 
$ E_{2g} $ $ {\bf \Gamma} $-phonons as well as $ A_{1}' $ $ {\bf K} $-phonons  
where we used  now 
the exact electron-phonon 
transition matrix elements in our numerics shown at the left hand side in 
(\ref{340}) and (\ref{350}) for the ${\bf \Gamma} $-phonons.   
We used here parameter values given in Refs.~\cite{Lazzeri1, Lazzeri2, Dietel3} 
and further that the phonon velocities of the transversal 
$ {\bf \Gamma} $-phonons 
is minus half of the velocity of the longitudinal ones \cite{Maultzsch1}.

\begin{figure}
\begin{center}
\includegraphics[clip,height=7.5cm,width=8.5cm]{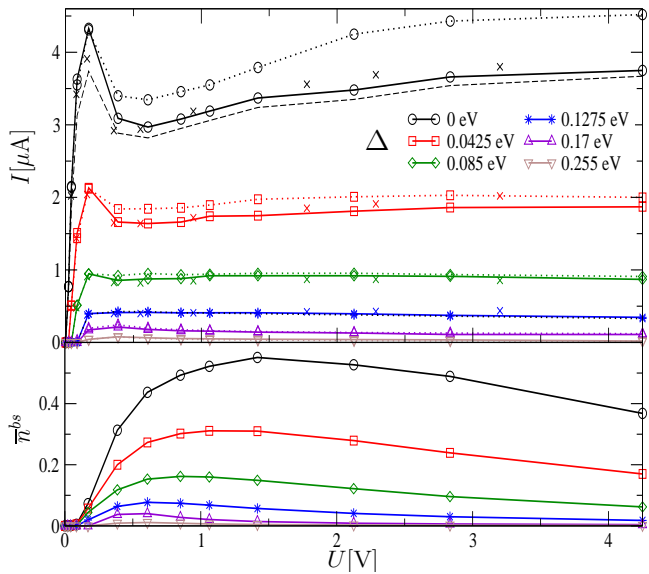}
\caption{(Color online) 
Upper panel with the continuous curves 
shows the current-voltage curve obtained using (\ref{10})-(\ref{20}) 
for a nanotube of length 
$ L= 300$\,nm, $l_e = 1600 $\,nm  \cite{Park1,Lazzeri2,Dietel3} 
and various gap energies $ \Delta $. The dashed curve is calculated 
for $ \Delta \to 0 $ and $ l_e = 800 $nm. The crosses uses the full two 
electron band, three phonon type theory with exact transition 
matrix elements. The dotted curves uses GW parameter values 
within the one-valley model for the electron-optical phonon scattering 
time \cite{Lazzeri3}. The lower panel shows 
the corresponding momentum and position averaged backward scattering 
phonon distribution function $ \overline{n}^{bs} $ defined in 
Ref.~\onlinecite{Dietel3}.}
\end{center}
\end{figure}

In the lower panel in Fig.~2 we show the 
corresponding energy-position averaged backward phonon distribution function 
$ \overline{n}^{bs} $ \cite{Dietel3} (\ref{10})-(\ref{40}).
We obtain that the phonon distribution function is 
around three to four times smaller 
for a SWNT after gap opening at $1.5$V 
when comparing it with the phonon distribution function values 
of the corresponding metallic nanotube \cite{Lazzeri1, Dietel3}.

\begin{figure}
\begin{center}
\includegraphics[clip,height=7.5cm,width=8.5cm]{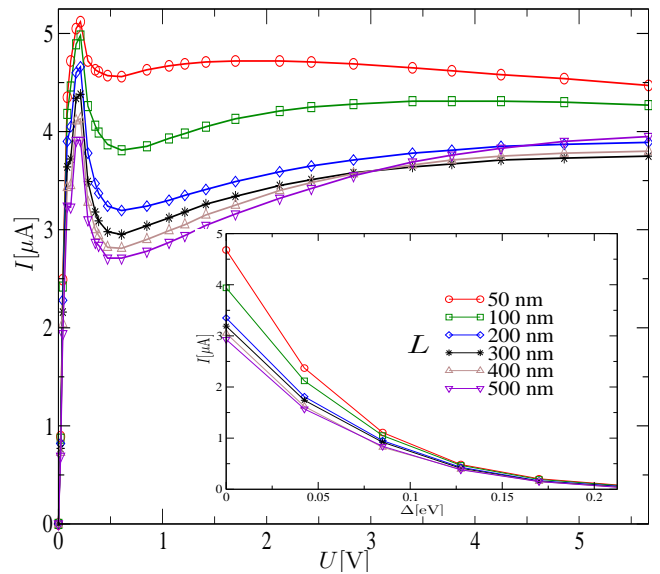}
\caption{(Color online)
We show the current-voltage curve for SWNTs with a gap 
$ \Delta  \to 0 $ and certain SWNT lengths $ L $ ($l_e =1600$nm).
The inset shows the current curve for a voltage 
$ U=1 V $ and several nanotube lengths $ L $ as a function of the gap energy 
$ \Delta $.}
\end{center}
\end{figure}

Figure~3 shows the current-voltage characteristic of a metallic SWNT  
where just the gap is opened, i.e. $ \Delta \to 0 $, for various 
nanotube lengths. Here we recorded more 
data points.
We obtain a dip in the current-voltage characteristic 
with a large negative differential 
conductance for $ U \gtrsim  \hbar \omega/e$ 
by taking into account that $ \hbar \omega \approx 0.17 $eV.
Fig.~2 shows that this behavior 
survives even at much smaller elastic scattering lengths. 
We can understand  this dip in the current curve in the following way: 
For small voltages $ k_B T/e \ll U \ll \hbar \omega/e $, 
phonon scattering can be neglected and 
the total current is the sum 
of the currents in each band separately. This current is for general 
$ \Delta $ approximately given by
\begin{equation} 
I_0 = 
\!\frac{8 e}{h} \!  \! 
\int\limits^{\infty}_{\Delta/2} \! \! d \epsilon n_F(\epsilon) = 
 \frac{8 e}{h} k_B T \left\{ \! \ln\left[ \! 1+\exp\left(\frac{\tilde{\Delta} }{2}
\right) \! \right]\! - \! \frac{\tilde{\Delta} }{2} \right\}    .      
     \label{50} 
\end{equation}
with $ \tilde{\Delta} \equiv \Delta/k_B T $ and  
$n_F $ is the Fermi-function.   
At room temperature $ T \approx 300 $K  
and small gaps $ \Delta \to 0 $
we have $ I_0 \approx 5.3 \mu $A.     
We point out that $ I_{0} $ is independent of the length 
$L $ of the SWNT. In (\ref{50}) we neglect elastic scattering,  
which is only a small correction factor since $ L/l_{e} \ll 1 $
for small SWNTs \cite{Dietel3}.
In the following discussion it is useful to have in mind  
that the energy broadening $ K_B T $ of the electrons 
(holes) in the conduction  (valence) band is much smaller than
the phonon frequency at room temperature, 
i.e. $ k_B T / \hbar \omega \sim 0.14 $.        
At higher voltages  $ U \gtrsim  \hbar \omega/ e $ first 
interband scattering starts. Due to the 
electron distribution function factors in 
the phonon scattering terms in (\ref{30}) and (\ref{40}) interband 
backward scattering from the conduction band to the valence  
band is then the dominant scattering contribution. This leads to a reduction 
of the current $ I_0 $ of order 
$ I^{U \sim \omega}_r \sim (e/h) k_B T  L/l_{\rm ep}  $ where the 
electron-phonon mean free path is given by 
$ l_{\rm ep}= v_F \tau_{\rm ep} \approx 130 $nm 
\cite{Dietel3}. At even higher voltages 
$ U \gg \hbar  \omega/ e $ also innerband scattering starts.
This leads to an energy diffusion of the electrons (holes) in the 
conduction (valence) band with an energy broadening from 
$ \Delta \epsilon \sim k_B T $ without scattering to  
$  \Delta \epsilon \sim (\hbar \omega/\pi)(1 +L/l_{\rm ep})$ 
for 
$ e U \gtrsim (\hbar \omega/\pi)(1 +L/l_{\rm ep})$ \cite{Dietel3}.
This broader  energy band consists of equally spaced subbands 
with separation distance $ \hbar \omega $ and 
energy width $ \sim k_B T $. This follows from  
$ k_B T \ll \hbar \omega $ and by taking 
into account the energy conservation in the electron-phonon scattering 
process.         
In this voltage regime the reduction of $ I_0 $ due to interband 
scattering is changed to 
$ I^{U \gg \omega}_r \sim (e/h) 
(k_B T /\pi) (1 +L/l_{\rm ep}) (\hbar \omega/e U) 
 L/l_{\rm ep} $. This leads to a vanishing of interband scattering at high 
voltages shown in Fig.~3 as a dip in the current-voltage characteristic 
with negative differential conductance.
From Fig.~3 we obtain that the considerations above are true only 
for $ L \ll 500 $nm. For $ L \approx 500 $nm we obtain a 
saturation of the depth and the width of the voltage dip. 
The reason lies in the fact that at larger nanotube lengths also 
secondary scattering processes like interband scattering from the valence band 
to the conduction band has to be taken into account which was not done  
in the scaling considerations above.  Also the phonon 
density dependence of the 
effective electron-phonon  scattering length  
$ l_{\rm ep} $ \cite{Dietel3} should be included.

At even higher voltages
we obtain from Fig.~3 that 
$ \lim_{U \to \infty} I \approx I_0 $ (apart from elastic scattering). 
This shows that a current reduction due to innerband 
scattering is very small in this regime.
Phonons which are created by electron scattering in one band 
can only scatter electrons in the other band on a length  
$ \Delta x \sim L (1 +L/l_{\rm ep})(\hbar \omega/eU)/\pi $ so that 
we can consider  
 the Boltzmann system (\ref{10})-(\ref{40}) in each band separately
at high voltages.
This leads to the result that 
the electrons (holes) are effectively down (up) scattered in the  
conduction (valence)  band. Then, by taking into account 
the fact of existing  band edges near the Fermi-niveau for small stresses,  
we obtain that innerband scattering does not influence the absolute current 
at high voltages.              

The inset in 
 Fig.~3 shows the current-voltage curve for a SWNT at
bias voltage $ U= 1 $\,V for various nanotube lengths as a function of the 
gap energy $ \Delta $. The functional behavior 
of the reduction factor from the $ \Delta \to 0 $ 
current value is reproduced well by  Eq. (\ref{50}).

Recently, it was shown for graphene \cite{Basko1, Lazzeri3} that 
screening effects in the electron-electron interaction 
are relevant for the calculation of the zone-boundary 
electron-phonon scattering time. 
In Ref.~\onlinecite{Lazzeri3}  
a numerical Green's function method (GW) was used in contrast to the standard 
density functional (DFT) theory in order to calculate 
this scattering time. Note that our one-valley model used so far 
is based on scattering parameters obtained by the latter calculation.  
The calculated values within the GW method are in better agreement with 
the experimentally obtained energy dispersion 
of the zone-boundary phonons 
for graphene than the values of the DFT calculation. 
For the zone-center phonons the agreement is less satisfactory with the 
experiments than in the DFT-method \cite{Piscanec1, Lazzeri4}.        
To our knowledge 
up to now in contrast to the DFT calculations for SWNTs 
in Ref.~\cite{Lazzeri1}  
an explicit  GW calculation for the electron-phonon scattering time in 
carbon nanotubes is still missing. Since electron screening is highly 
dimension dependent this calculation would be of course of general interest. 
Nevertheless we are now able to calculate from 
Ref.~\cite{Lazzeri3} within the GW method
by using the zone-folding method \cite{Lazzeri1} effective 
electron-phonon scattering times for zone-center and zone-boundary 
phonon scattering for SWNTs. By  using Mathiessen rule 
we obtain an effective electron-phonon scattering length 
 $ l_{\rm ep}= 68 $nm for an SWNT with diameter 
$ 2 $nm. Corresponding scattering times for the phonons due to 
electron-phonon scattering, the averaged phonon frequencies and 
phonon velocity for the one-valley model can be calculated accordingly 
by using the definitions in Ref.~\onlinecite{Dietel3}. 
The acoustic-optical phonon 
scattering ratio $ \tau_{\rm ac}/p\tau_{\rm op} $ \cite{Dietel3} is then 
determined from the theoretical determined current-voltage characteristic 
of the full two-valley model by comparison with the experimentally given 
current-voltage characteristic for the $ L=300$nm SWNT without stress. 
We obtain values 
$ \tau_{\rm ac}/p\tau_{\rm op} = 3.5$ for an optimal accordance to the 
differential conductance at high voltages. The absolute theoretical 
current-value is then around $ 20 $\% lower in the high voltage regime 
than in the experiment. By using these theoretical GW quantities     
we show by the dotted curves 
in the upper panel in Fig.~1 the current-voltage characteristic 
for a $ L=300 $nm SWNT and various gap energies $ \Delta $ within an 
effective one-valley model.     
    
Up to now we only considered metallic nanotubes.
Semiconducting nanotubes show already a gap $ \Delta $
without stress. Since  $ \Delta \gg \hbar \omega $ for small  diameter 
semiconducting SWNTs ($\sim 2 $nm) we obtain a 
vanishingly small current-voltage curve in this case. For semiconducting 
tubes with much larger diameters, the energy gap $ \Delta $ 
is decreasing but in order to obtain quantitatively the current-voltage 
characteristic one should  
take into account also the  much larger 
electron-phonon scattering time $ \tau_{\rm ep} $ \cite{Lazzeri1} 
of such systems and further 
that now also higher transversal electron bands become important. 
Nevertheless, we expect from the discussion above a qualitative similar 
current-voltage behavior, with a negative differential conductance at 
$ U \gtrsim \hbar \omega/e $  as for the metallic nanotubes. 
Similar should be true for graphene nanoribbons. 
Remember that the existence  
of a negative differential conductance region in the current-voltage 
characteristic is mainly based on the gap opening under stress and 
that we have $ k_B T \ll \hbar \omega $ at room temperature.   
Here we note that corresponding to armchair nanotubes discussed above 
metallic zigzag-edged  nanoribbons do not show a gap opening under axial 
stress in contrast to armchair edged ones. This is shown by numerical 
calculations in Ref.~\onlinecite{Poetschke1} based on an atomistic model. 
This follows also immediately from 
the electron-phonon interaction Hamiltonian for
acoustic phonons in Ref.~\onlinecite{Suzuura1} which we 
used above in the derivation  of (\ref{300}) and (\ref{310}).

Finally, we briefly discuss the current-voltage characteristic of  a SWNT in 
the superelongation regime. In Ref.~\onlinecite{Dietel1}
 we obtain that with increasing 
external stress defect stripes are created  with increasing lengths 
for increasing stress. This is seen by a kink motion
at the boundary surface of the SWNT 
in transmission electron microscope pictures \cite{Huang1}. 
The number of these stripes 
is proportional to the external stress 
whose lengths corresponds to missing lattice plane pieces in the crystal. 
In first approximation we expect that the current is given  
by the electrons running over complete lattice planes. The 
electrons which run over incomplete lattice planes 
are completely backscattered and these do not 
contribute to the current. For those electrons which run 
over the complete lattice planes we further have to consider the linear 
elasticity current corrections which we have discussed at length above. 
All this means that for armchair SWNTs, which do not 
show a stress dependent gap we expect a linear decreasing 
current curve as a function of the external stress. Such a behavior is in 
fact experimentally observed in Ref.~\onlinecite{Huang1}.

\section{Conclusion}       
We have calculated
the current-voltage characteristic of  
metallic carbon nanotubes under homogeneous axial stress in the 
linear elastic regime lying on a substrate by means  
of a semi-empirical Boltzmann equation for 
electrons and phonons. We have found    
that the stress leads to an  energy gap which, 
causes a large negative differential conductance at 
$ U \gtrsim \hbar \omega/e $. 
Moreover this negative differential conductance is largest  
at small stress just after the gap is opened,
getting smaller with a larger stress applied.  
Materials with such a behavior are very useful 
for the activation and 
deattenuation of oscillating circuits.
For larger voltages at $ U \approx 1.5 V $ and small stress
just after the gap opening 
we obtain approximately six times smaller current values for 
the $L=300 $nm nanotube  in comparison 
to the tube without stress. The corresponding momentum and position 
averaged optical phonon distribution function is around
three to four times smaller. In addition to the discussion of 
the current-voltage characteristic in the linear elastic regime 
we have also discussed modifications of the characteristic 
in the superelongation regime. 

Concerning the current-voltage characteristic of 
large diameter semiconducting nanotubes and graphene 
nanoribbon, we  have argumented that 
a qualitative similar differential negative behaviour 
as for the metallic nanotube 
should be seen.

\acknowledgments
The authors acknowledge the useful discussion 
with A.~Lima.   
We further acknowledge the support provided by Deutsche Forschungsgemeinschaft
under grant KL 256/42-3.


\begin{thebibliography}{99}
\bibitem{REMA}
 Our space grid  
is the same, while  
our 
momentum grid is twice as large as
in 
the stress-free calculations of 
Ref.~[\onlinecite{Dietel3}].
\end{thebibliography}
\end{document}